\documentclass[11pt, onecolumn, a4paper]{article}
\usepackage{amsmath}
\usepackage{amssymb}
\usepackage{amsthm}
\usepackage{lipsum}
\usepackage{mathptmx}

\usepackage[rm,up,sc,topmarks,calcwidth,pagestyles]{titlesec}
\titleformat{\section}{\Large\bfseries\rmfamily}{\thesection}{1em}{}
\titleformat{\subsection}{\large\bfseries\rmfamily}{\thesubsection}{1em}{}
\titleformat{\subsubsection}{\large\it\rmfamily}{\thesubsubsection}{1em}{}

\usepackage{cite}
\usepackage{algorithmic}
\usepackage[dvipdfmx]{graphicx}
\usepackage{textcomp}
\usepackage{url}

\usepackage{color}

\title{Key Factor Not to Drop Out is to Attend the Lecture}
\author{Hideo Hirose \\
Hiroshima Institute of Technology, Hiroshima, Japan}
\date{}
\begin{document}
\maketitle
\thispagestyle{empty}

\section*{Abstract}
In addition to the learning check testing results performed at each lectures, we have extended the factors to find the key dropping out factors. Among them are, the number of successes in the learning check testing, the number of attendances to the follow-up program classes, and etc.
Then, we have found key factors strongly related to the students at risk. 
They are the following. 
1) Badly failed students (score range is 0-39 in the final examination) tend to be absent for the regular classes and fail in the learning check testing even if they attended, and they are very reluctant to attend the follow-up program classes. 2) Successful students (score range is 60-100 in the final examination) attend classes and get good scores in every learning check testing. 3) Failed students but not so badly (score range is 40-59 in the final examination) reveal both sides of features appeared in score range of 0-39 and score range of 60-100. Therefore, it is crucial to attend the lectures in order not to drop out.
Students who failed in learning check testing more than half out of all testing times almost absolutely failed in the final examination, which could cause the drop out. Also, students who were successful to learning check testing more than two third out of all testing times took better score in the final examination.
%
%
\\[2mm]
{\it Keywords: }learning check testing, placement test, follow-up program, item response theory, multiple linear regression, final examination.








\section{Introduction}

It is crucial to identify students at risk for failing courses and/or dropping out as early as possible because a variety of students are now enrolled in universities and we teachers have to educate them altogether.
This circumstance prohibits us to use conventional methods such as a mass education method. However, the number of staffs and classes are limited. New assisting systems using ICTs shall be introduced to solve such a difficulty. 
To overcome this, we established online testing systems aimed at helping students who need further learning skills in mathematics subjects.
Such systems include 1) the learning check testing, the LCT, for every class to check if students comprehend the contents of lectures or not, 2) the collaborative working testing, the CWT, for training skills with supporters and teachers, and 3) the follow-up program testing, the FPT, to check if the follow-up program class members understand the standard level of the lectures.
The system has been successfully operating (see \cite{LTLE2016a}, \cite{LTLE2016b}), and some computational results were reported \cite{LTLE2017}. In addition, other relevant cases were well investigated (see \cite{LTLE2016c}, \cite{BIC2016}, \cite{IEE2018}, \cite{PISM2018}, \cite{IJSCAI2017}, \cite{LTLE2016d}). 

Using the accumulated data in the database, we may find some key factors strongly related to the students at risk,
as indicated in \cite{Elouazizi}, \cite{Siemens2015}, \cite{Siemens2012}, and \cite{Waddington2016}, if we pay attention to learning analytics.
Then, we may be able to actively make a appropriate decision for better learning methods. 
As indicated in \cite{WiseShaffer}, it is also important to analyze the data theoretically. 

This paper is aimed at obtaining effective learning strategies for students at risk for failing courses and/or dropping out, using a large-scale of learning data accumulated from the follow-up program systems.
They consists of the placement scores, every LCT scores, FPT success/failure times, FPC attendances, etc.
In this paper, we use the ability values for students' learning skills obtained from the item response theory (IRT, e.g., see \cite{Ayala}, \cite{Hambleton91}, \cite{LindenHambleton}).
Although the subjects we deal with are analysis basic (similar to calculus) and linear algebra, we show the case of linear algebra as a typical case.

\section{Success/Failure Responses and the LCT Ability Values}

The LCT is a kind of short-time testing using five questions in each LCT in the first semester in 2017. All the students in regular classes take the LCT for ten minutes via the online testing system. All the questions are the same to each student, but the order to each question to a student is sorted in a different order from the next student. We have fourteen lectures with one midterm and one final examinations in the semester; thus, the number of LCT is fourteen. We can estimate the students abilities to each LCT using the item response theory (IRT) evaluation method. 

Each problem in five items consists of multiple small questions. Students select appropriate answers to each small question from many choices. 
We adopts the two-parameter logistic function $P(\theta_i;a_j,b_j)$ shown below instead of the three-parameter logistic function including pseudo-guessing parameter. 
\begin{eqnarray} 
P_{i,j}=P(\theta_i;a_j,b_j)={1 \over 1+\exp\{-1.7a_j(\theta_i-b_j)\} }=1-Q_{i,j},
\end{eqnarray}
where $\theta_i$ expresses the ability for student $i$, and $a_j, b_j$ are constants in the logistic function for item $j$, and they are called the discrimination parameter and the difficulty parameter, respectively. 
Then, the likelihood for all the examinees, $i=1,2,\dots,N$, and all the items, $j=1,2,\dots,n$, will become
\begin{eqnarray} 
L=\prod_{i=1}^N \prod_{j=1}^n \left(P_{i,j}^{\delta_{i,j}} 
\times Q_{i,j}^{1-\delta_{i,j}} \right),
\end{eqnarray}
where $\delta_{i,j}$ denotes the indicator function such that $\delta=1$ for success and $\delta=0$ for failure. 
In a sense, $P_{i,j}$ in Equation (1) is a logistic probability distribution function with unknown parameters $a_j$ and $b_j$, and the random variable is $\theta_i$. 
However, $a_j$, $b_j$, and $\theta_i$ are all unknown here. We have to obtain the maximum likelihood estimates for $a_j$ and $b_j$, and $\theta_i$ simultaneously by maximizing $L$ in Equation (2). 

However, as easily imagined with so small number of questions, the estimated ability values tend to have biases and the variances are large (see \cite{IJSCAI2017}, \cite{LTLE2016d}). It would be difficult to classify the students into a successful group and a failed group in the final examination using each LCT result. 
Thus, we first use all the LCT results in classifying.

Figure 1 shows the histogram of estimated abilities of LCT to successful students overlaid the histogram of estimated abilities of LCT to failed students in the case of linear algebra in the first semester in 2017. We can see that it would be difficult to find the optimal discriminating threshold to success/failure students.
The numbers of successful students is 898, and failed students is 145; the ratio of failed students to all the students is $0.14$.

\begin{figure}[htbp]
\begin{center}
\includegraphics[height=6.5cm]{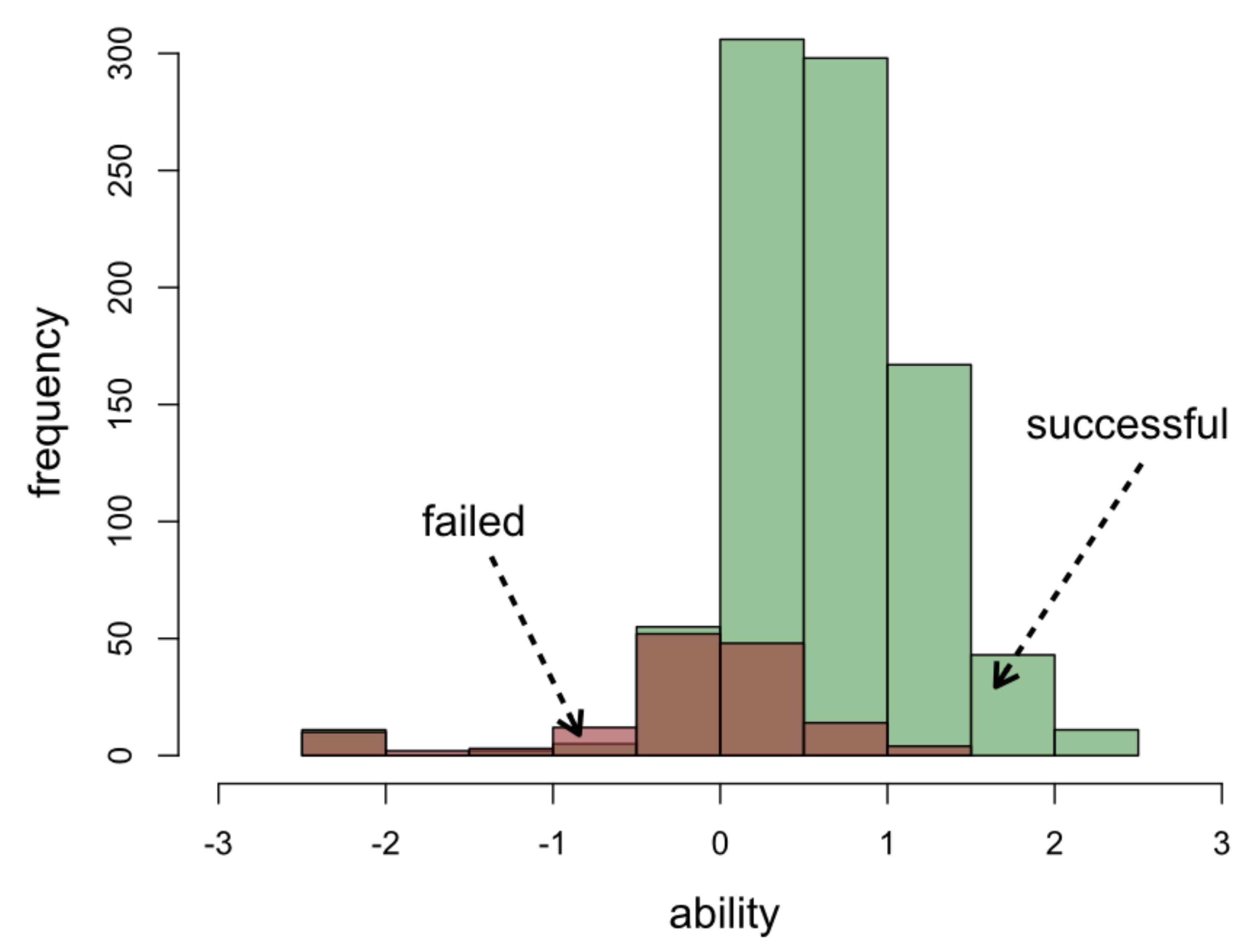}
\end{center}
\caption{Histograms of estimated abilities of LCT to successful students and to failed students (linear algebra in the first semester in 2017).}
\end{figure}

Except for very low values of estimates, the histograms indicate the normal distributions with different mean values (around $0.63$ for successful students and $-0.17$ for failed students); the lowest estimates around $-3.0$ in both groups were resulting from the absence for testing. However, it seems very difficult to discriminate students into two groups by using certain ability threshold value. When we adopt the decision tree method, the most appropriate ability threshold values becomes to be $-0.1065$.

The confusion matrix using this threshold is illustrated in table 1. The misclassification rate for this confusion matrix is $0.11$.  Limited to failed students, the decision tree predicted 107 students may fail, and 70 students actually failed; the hitting ratio is $65\%$.

\begin{table}[htbp]
\caption{Confusion matrix determined by decision tree using full response matrix.}
\begin{center}
\begin{tabular}{cc|ccc}
\hline
&&&predicted \\
 &  & successful & failed & total \\
\hline
                 & successful & 861 & 37& 898 \\
observed  & failed & 75 & 70 & 145\\
                 & total & 936 & 107 & 1043\\
\hline
\multicolumn{4}{l}{\qquad \qquad \qquad \qquad \quad threshold $= -0.1065$}
\end{tabular}
\label{tab1}
\end{center}
\end{table}

In addition to the LCT results, we have incorporated the placement test (PT) results taken at the very beginning of the first semester. We have two kinds of tests: one is rather fundamental test and the other is advanced test in high school level. Using the fundamental PT and the LCT results, we plotted the correlations for these two tests in three groups in Figure 2 in the case of linear algebra in the first semester in 2017; first group is the successful in the final examination (score range is 60-100 expressed by green dots in the figure), second group is the badly failed group (score range is 0-39 expressed by red dots), and the rest is the group (score range is 40-59 expressed by yellow dots). The horizontal axis means the ability values standardized to the standard normal distribution, and the vertical axis means the fundamental PT score. 
Although the information is added, it is still hard to find the boundaries to classify students into three groups or two successful/failed groups.
In order to discriminate the successful/failed students much more clearly, it would be recommended to include other kind of information.

\begin{figure}[htbp]
\begin{center}
\includegraphics[height=5.5cm]{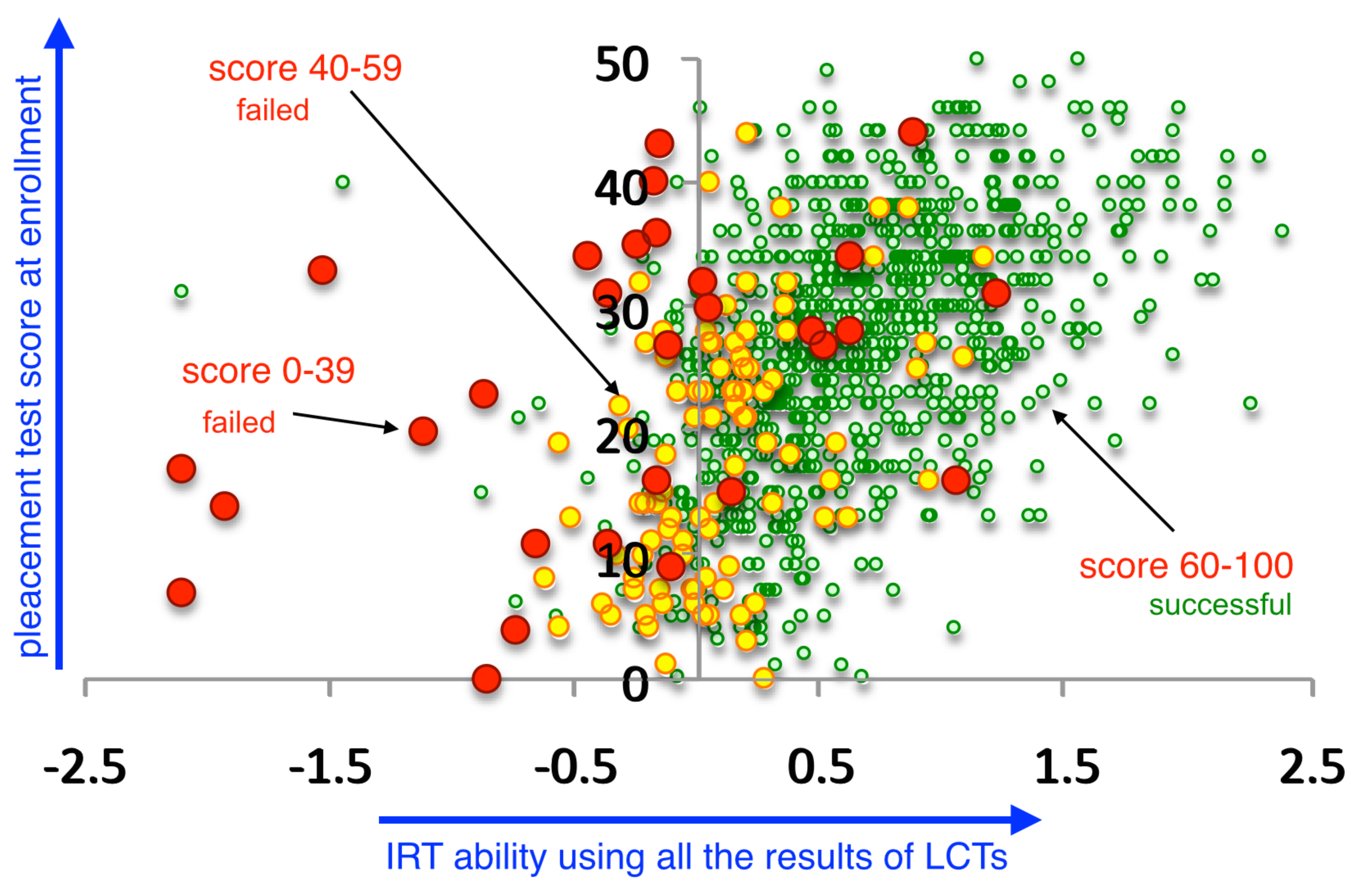}
\end{center}
\caption{Correlations for the LCT results and the placement test results in three successful/failed groups (linear algebra in the first semester in 2017).}
\end{figure}

\section{Attendance to the Lectures and the Follow-up Program Classes}


Attendance/absence to classes is other discrete type information. Intuitively, we feel that the more frequently attend the classes, the higher the scores of the final examination. Recently, it is often seen that attendance/absence information is memorized to the database automatically using the electric card attendance check system. However, the system is not perfectly working; some students may disappear after exposing their cards. 

The LCT compensate this defect. The attendance information cannot be guaranteed unless the testing is completed. Figure 3 shows that the attendance/absence information are classified into three groups: the first is for score range is 60-100 seen on the right in the figure, the second is for score range is 40-59 seen in the middle, and third is for score range is 0-39 seen on the left. In these matrices, row means the student id, and column means the question id. Using two kinds of attendance/absence information by electric cards (expressed by $y$ shown below) and LCT results (expressed by $x$ shown below), the value of each element, $s$, is determined and is colored by the formula of $s = 10 x + y$, where meanings of $s$, $x$, and $y$ are indicated in Figure 4. The figure shows the scheme of the attendance/absence information and LCT successful/failed information.
For example, $s=55$ means that a student was absolutely absent for the class, and $s=11$ means that a student is absolutely attended the class; they are also indicated in Figure 3.


Since each element is colored by green to red according to $s$ value from lower to higher, red and orange colors indicate the absence or failed in the LCT, and green color indicate the success in the LCT.
Obviously, three groups can be classified clearly by these colors by looking at the figure.
This indicates that the attendance/absence information may play a key role in determining the risk of a student in addition to the LCT results.

\begin{figure}[htbp]
\begin{center}
\includegraphics[height=9.5cm]{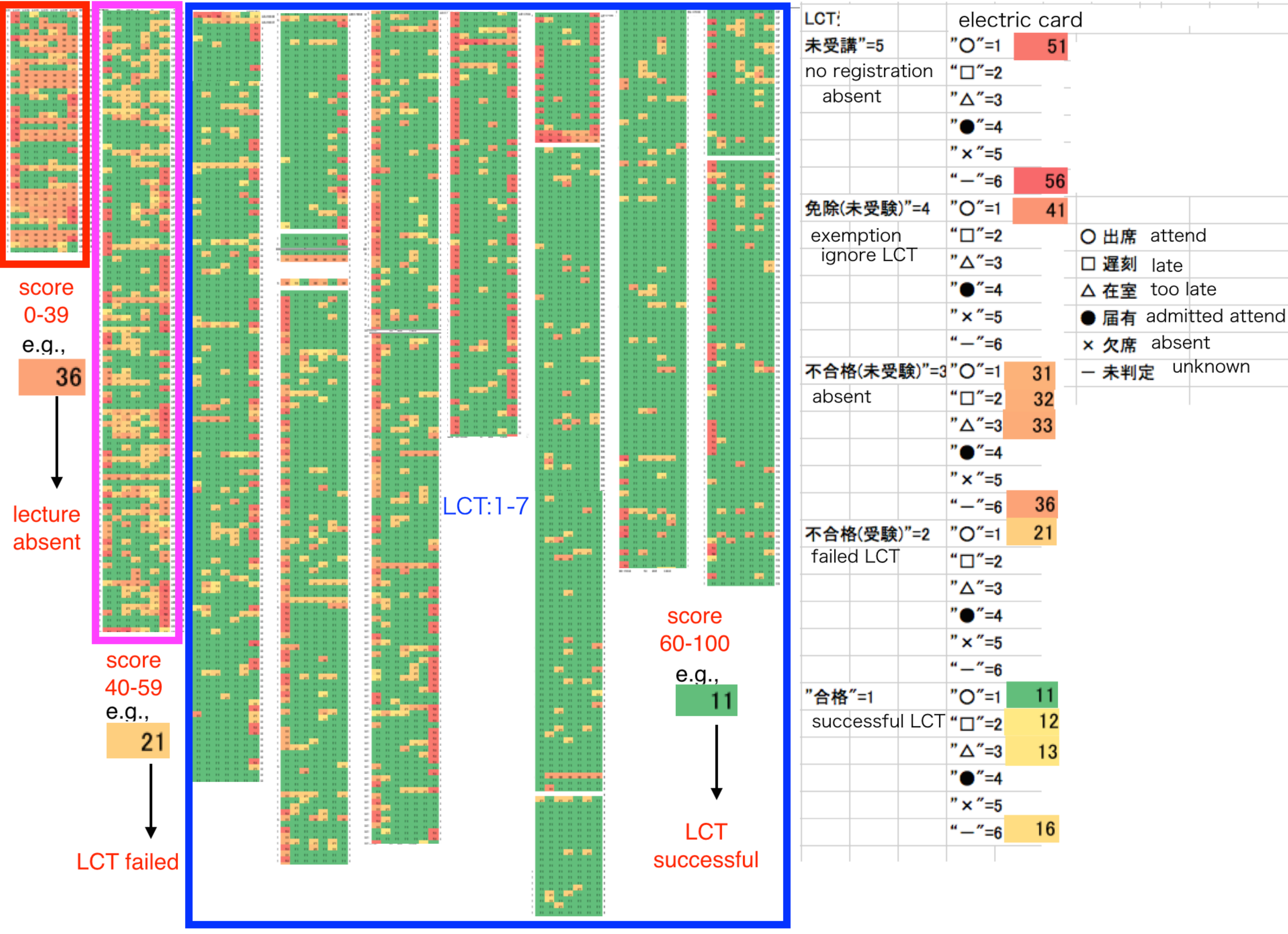}
\end{center}
\caption{Three groups classified by using the attendance/absence information and LCT successful/failed information.}
\end{figure}

\begin{figure}[htbp]
\begin{center}
\includegraphics[height=9cm]{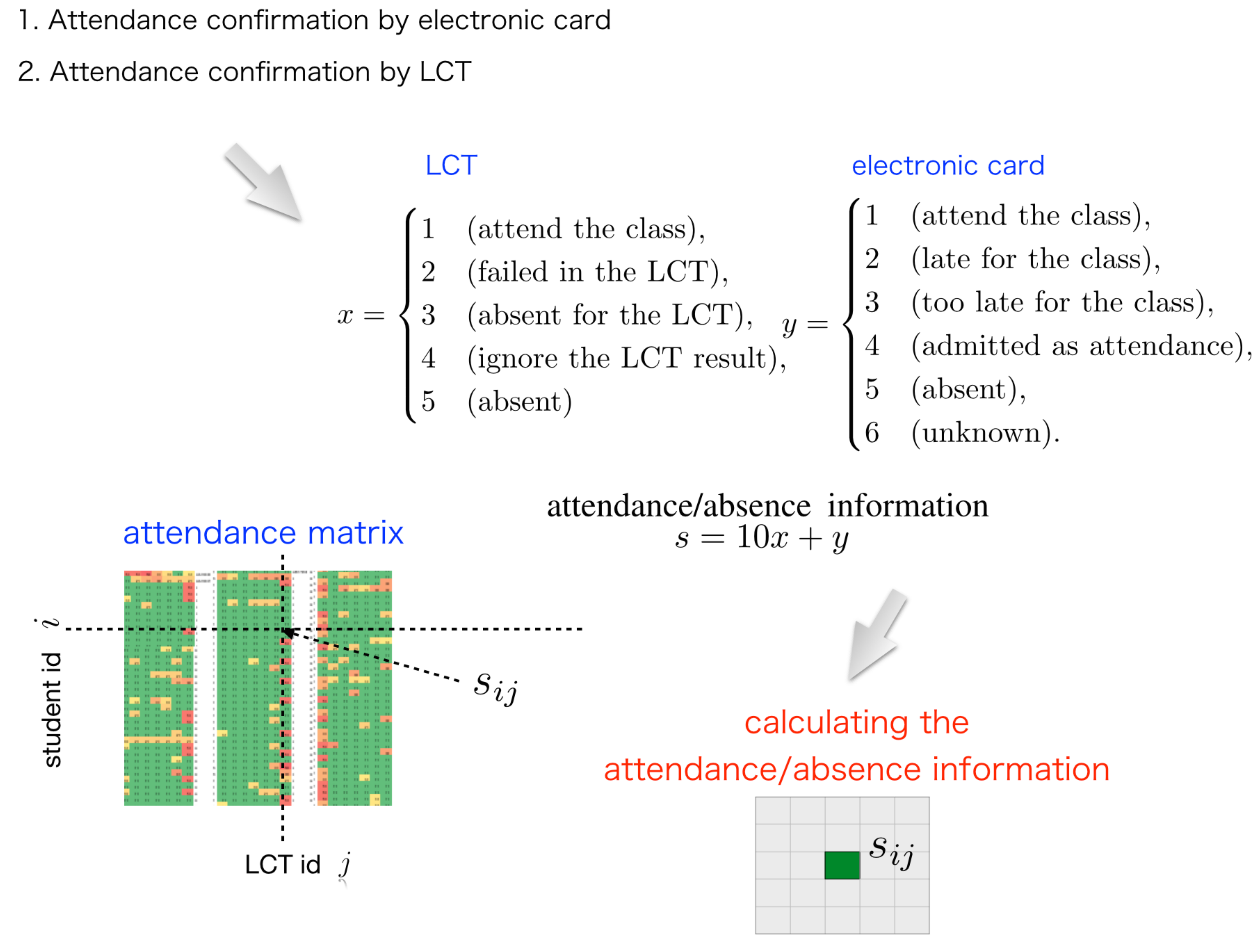}
\end{center}
\caption{Scheme of the attendance/absence information and LCT successful/failed information.}
\end{figure}

\section{Finding the Important Factors for Risk}

We first show the relationships among the factors we are concerned with in Figures 5 and 6.
In these figures, for example, we see that there is a strong relationship between the LCT successes and the no requirement for the FPT (see first column and sixth row in the figures), but it seems unclear which factors are key factors in classifying the successful/failed groups.
In this paper, however, we will not deeply discuss the dimension reduction problem. 
We are only interested in finding the key factors related to the risky students in the final examination.
Thus, a much easier method will be taken in the following.

\begin{figure}[htbp]
\begin{center}
\includegraphics[height=8cm]{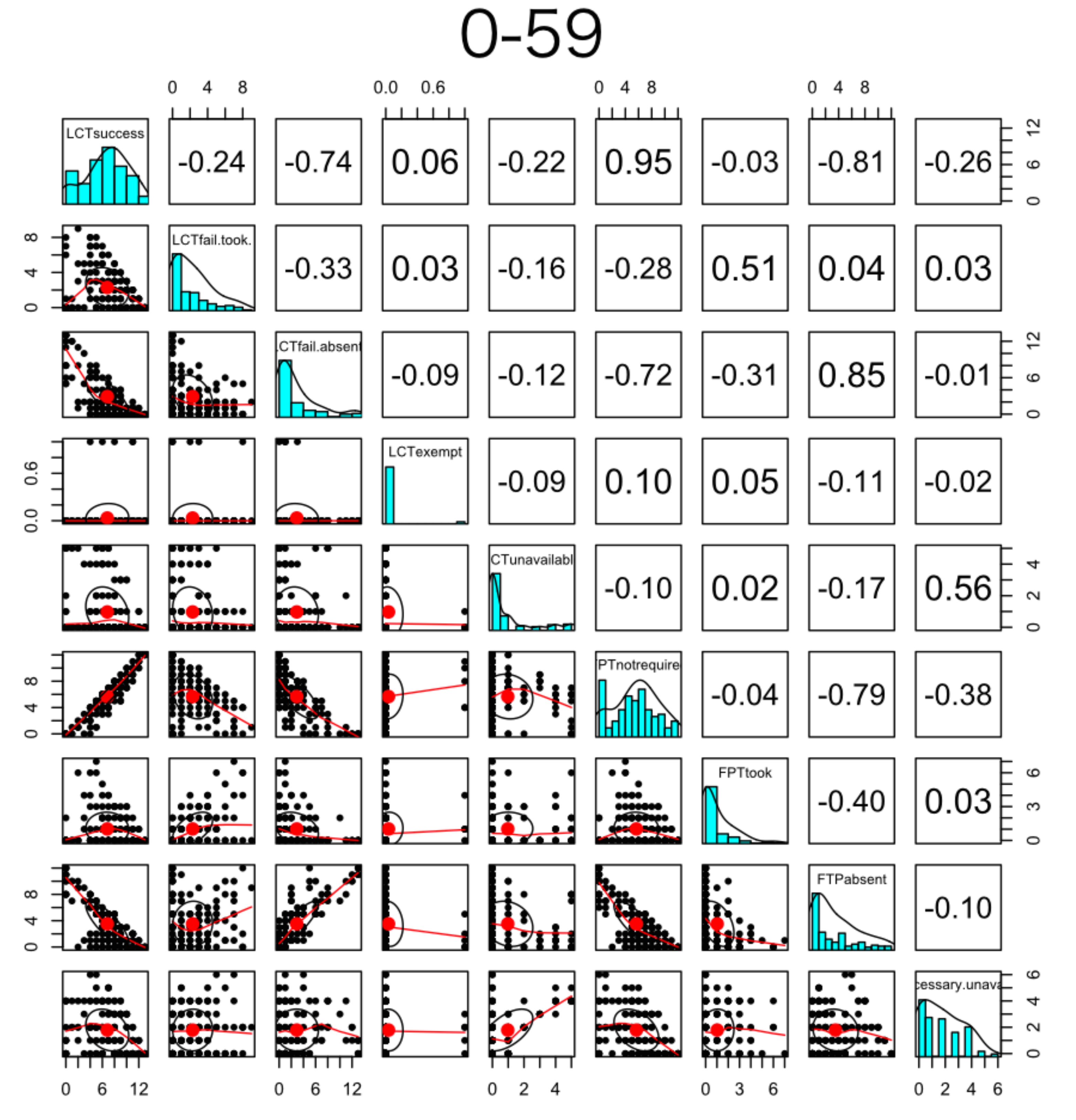}
\end{center}
\caption{Relationships among the factors when the score range is 0-59.}
\end{figure}
\begin{figure}[htbp]
\begin{center}
\includegraphics[height=8cm]{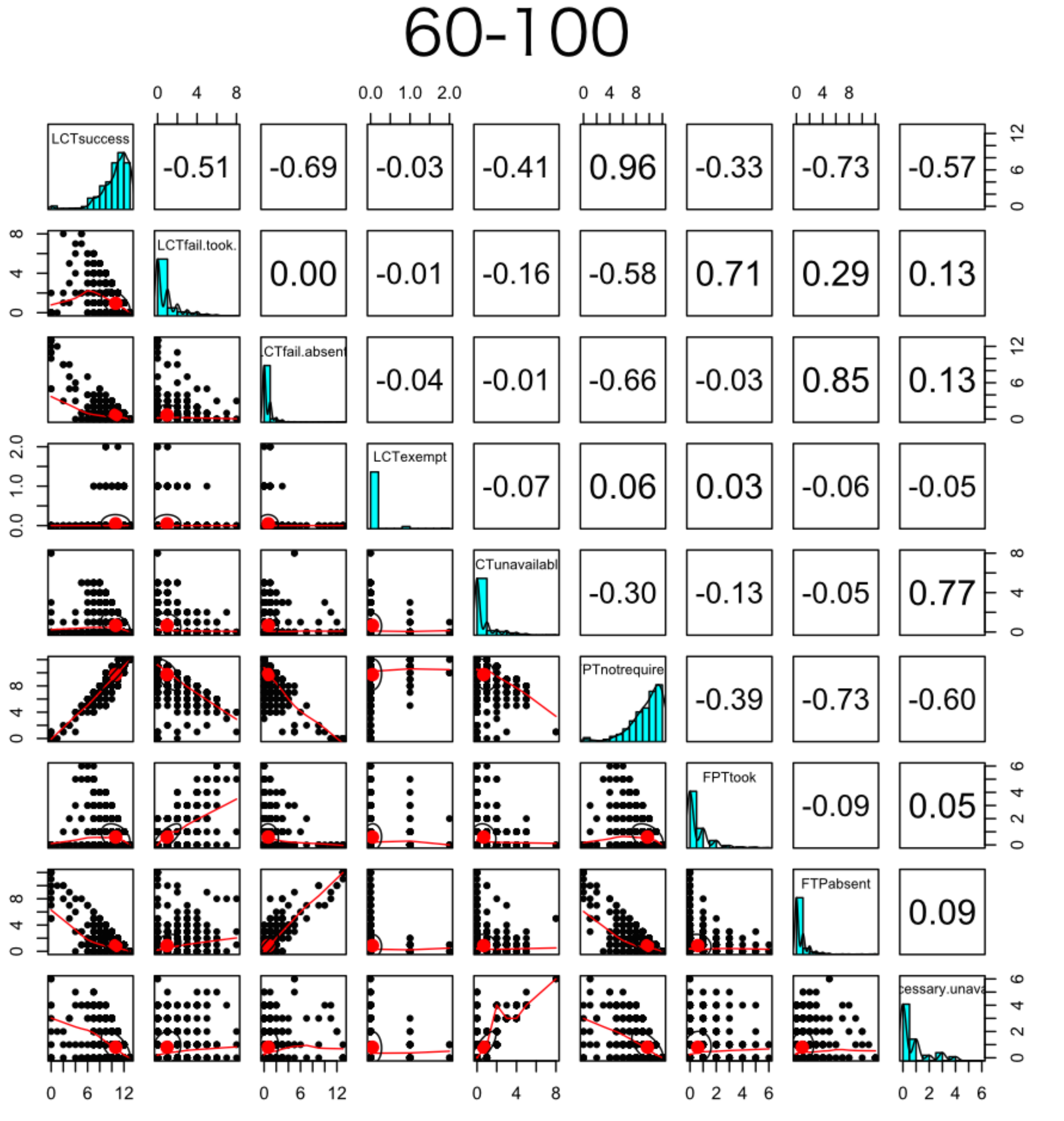}
\end{center}
\caption{Relationships among the factors when the score range is 60-100.}
\end{figure}

Since we have known that the attendance/absence information may be effective for classifying the students groups into successful/failed students in the final examination, we apply the multiple regression analysis of $Y=X\beta$ in finding the key factors; candidates of factors are shown in Figure 7.




\begin{figure}[htbp]
\begin{center}
\includegraphics[height=7cm]{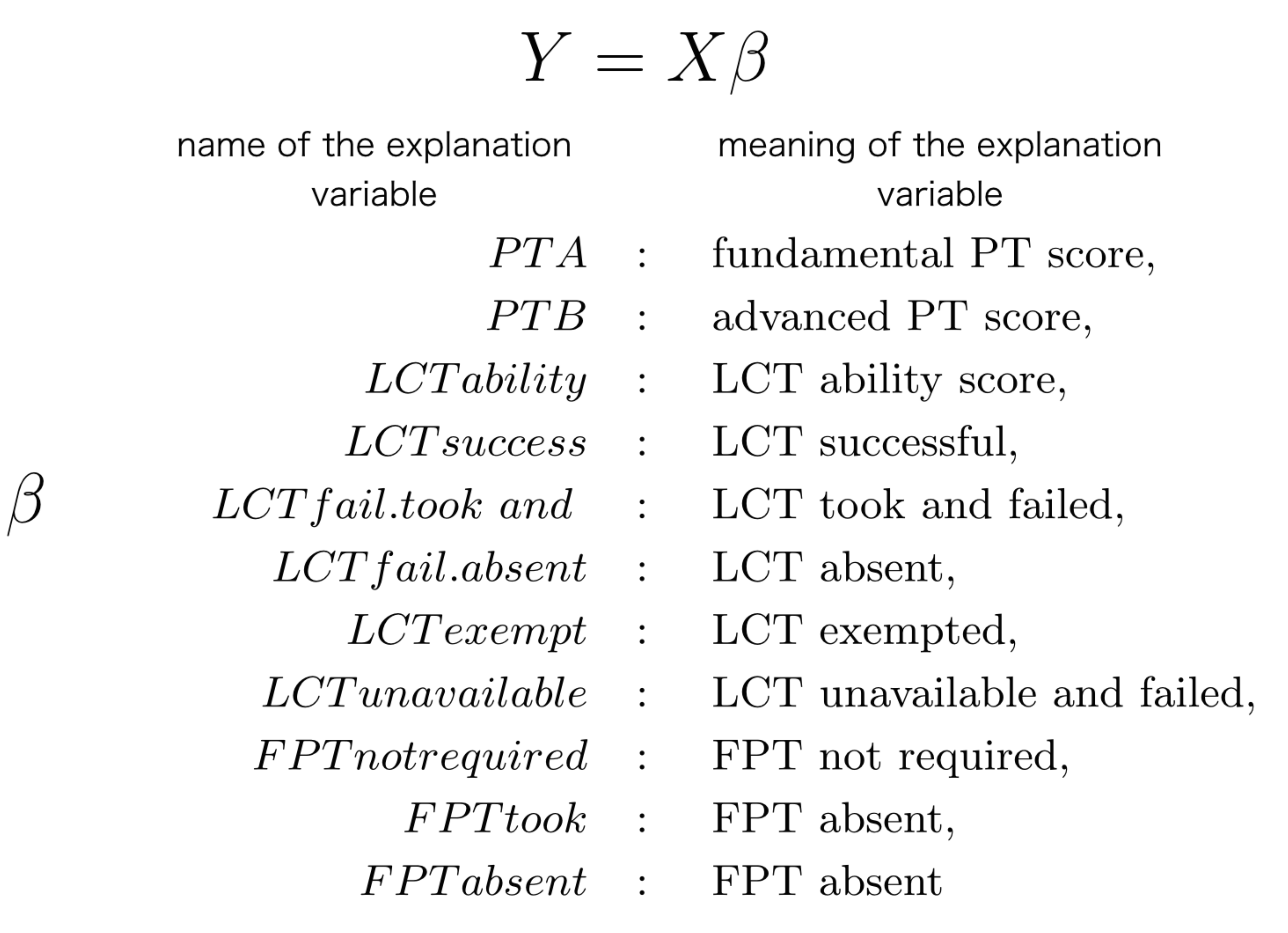}
\end{center}
\caption{Factors in the multiple regression analysis.}
\end{figure}

Applying the multiple linear regression using the accumulated learning data, e.g., estimated LCT ability values, placement scores, class attendance/absence, follow-up class attendance/absence, and etc., we obtained the result shown in Figure 8.
Marked symbols by asterisks indicate that these factors are significant with given $p$-values in using R {\cite R}, the statistical computing and graphics language and environment. The symbol of FPTnotrequired means that students took the LCT and successful, resulting no requirement for follow-up class attendance. That is, attendance/absence for FPT is the most significant information in deciding successful/failed students.

\begin{figure}[htbp]
\begin{center}
\includegraphics[height=4.3cm]{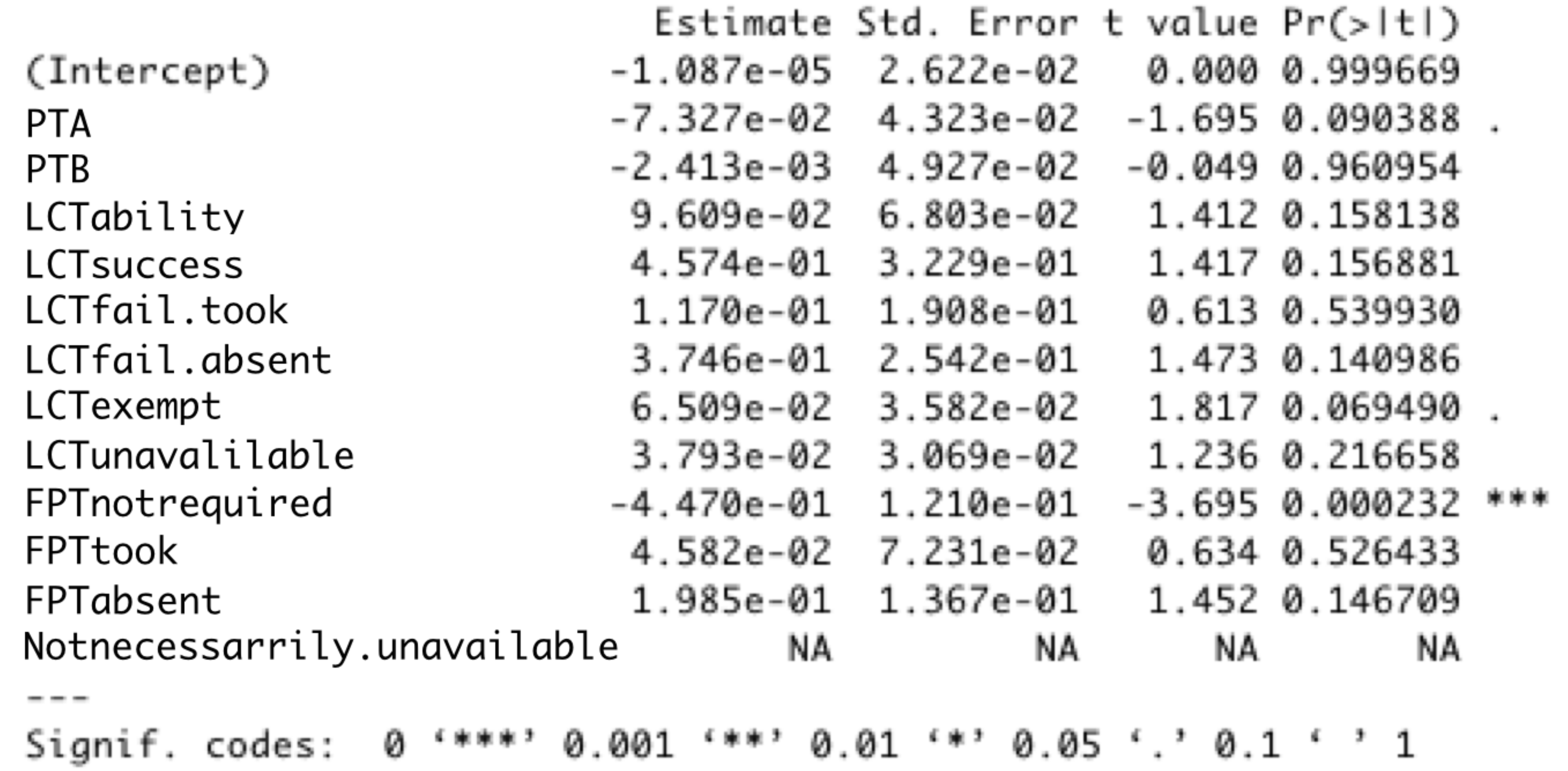}
\end{center}
\caption{Multiple linear regression analysis result.}
\end{figure}

Therefore, we next focus on this factor. Figure 9 shows the 2-dimensional relationship between the number of successes in the LCT and the number of absents for the follow-up classes for the three groups, score ranges are 60-100, 40-59, and 0-39 in the final examination. 
At a first glance, we can see that a clear linear relationship between the number of successes in the LCT and the number of absents for the FPC when score range is 0-39.
We also see some similarity between the cases score range 40-59 and the cases score range 60-100, but it is unclear.
Since each dot represents a student in the figure, overlaid dots representing the same position hinder the accumulated numbers of the students. 

\begin{figure}[htbp]
\begin{center}
\includegraphics[height=8cm]{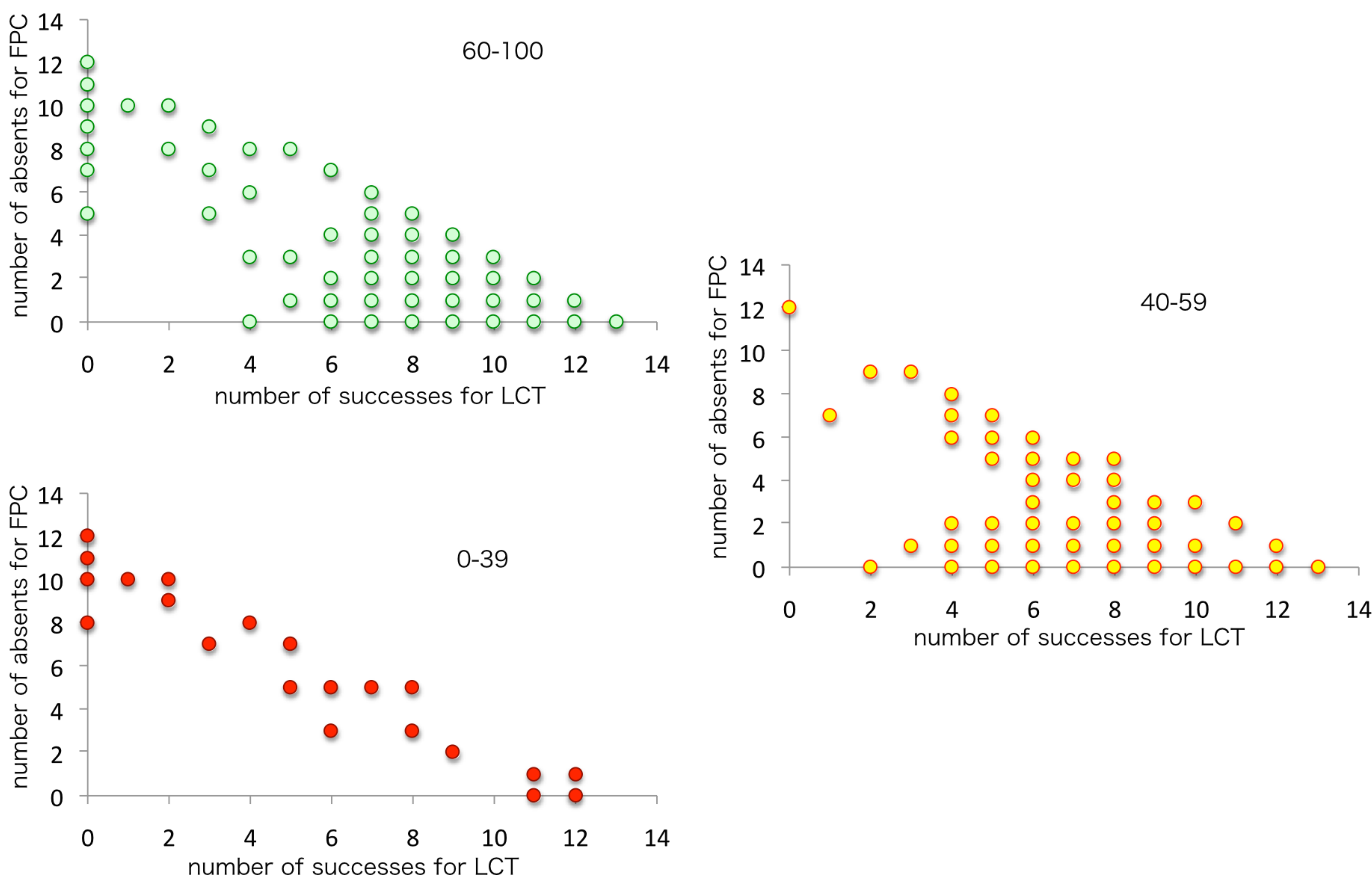}
\end{center}
\caption{2-dimensional relationship between the number of successes in the LCT and the number of absents for the FPC.}
\end{figure}

Figure 10 shows the 3-dimensional bar charts representing the relationship between the number of successes in the LCT and the number of absents for the follow-up classes for the three groups, score ranges are 60-100, 40-59, and 0-39 in the final examination.
By looking at the figure, we find the following: 1) When score range is 60-100, almost all the students show successful results in the LCT and very small number of absences for the FPC (almost all are not required the attendance for the FPC). 2) When score range is 0-39, we see a clear linear relationship between the number of successes in the LCT and the number of absents for the FPC, which means that almost all the failed students in the LCT or students absent for the classes ignore the attendance for the FPC. 3) When score range is 40-59, students reveal both sides of features appeared in score range of 0-39 and score range of 60-100. Some students tried to make effort to be successful, and some were successful but unfortunately some were not. Therefore, we have found that failed students in the final examination were reluctant to attend the classes and showed failed LCT results, and they were unwilling to attend the FPC in addition. As intuition suggests, the most crucial factor for the success in the final examination is attendance to the class.

\begin{figure}[htbp]
\begin{center}
\includegraphics[height=8cm]{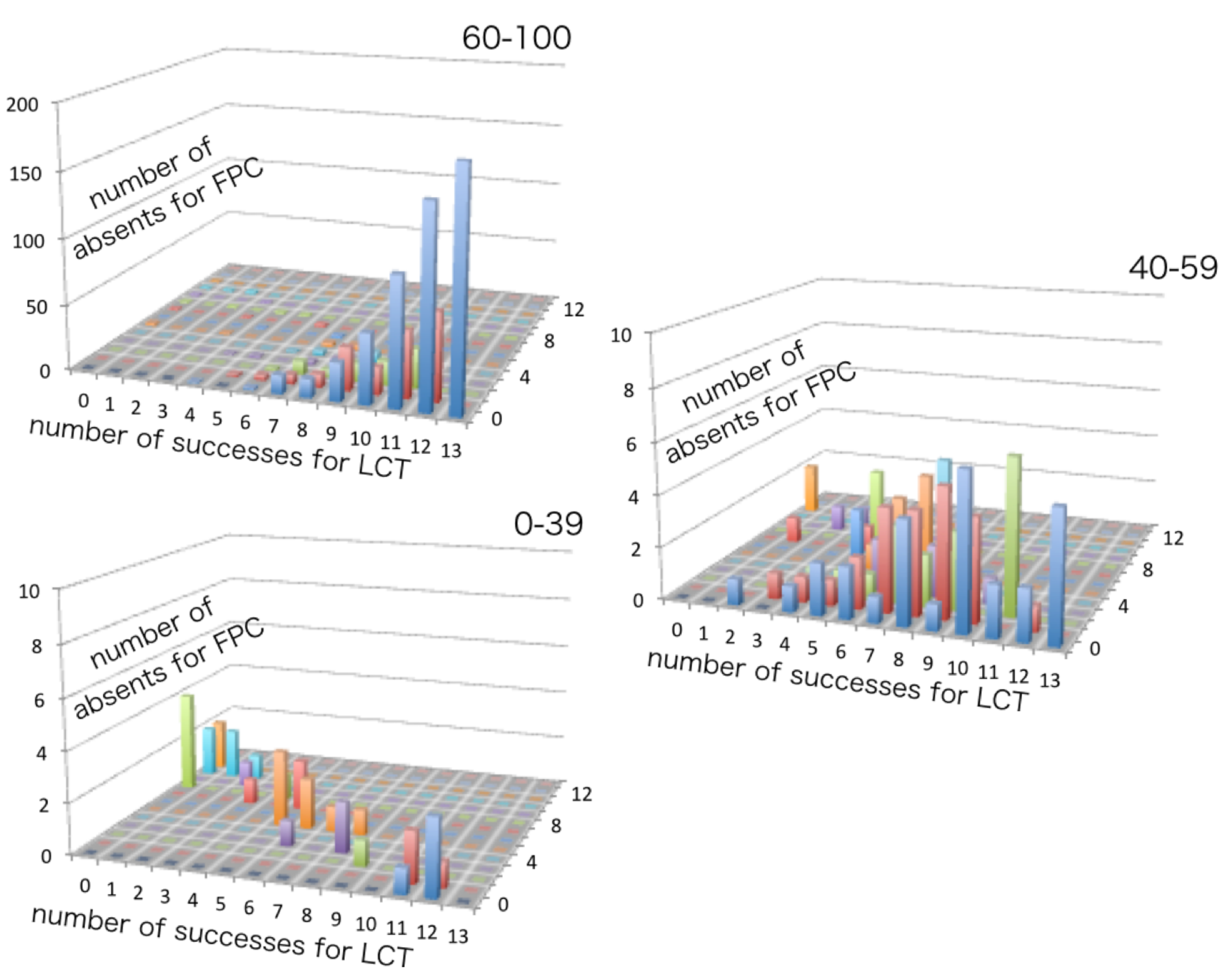}
\end{center}
\caption{3-dimensional bar charts representing the relationship between the number of successes in the LCT and the number of absents for the FPC.}
\end{figure}

\section{Discussions}


We have been looking at some factors to classify successes and failures in the final examination. To investigate such factors much more precisely, more detailed information may be required. Thus, we have classified the successful group into four groups such as A+, A, B, C, where scores in these groups are distributed to be  90-100, 80-89, 70-79, 60-69. The possible factor to discriminate these groups is considered to be the number of successful LCT. 

Figure 11 shows the frequency bar charts for the number of successful LCT to each group. 
Taking a look at the figure, we can see that students who failed in LCT more than seven times almost absolutely failed in the final examination, which could cause the drop out. Also, students who were successful to LCT more than ten times took better score in the final examination. Since all the testing times were 13 in this case, this means that students who failed in learning check testing more than half out of all testing times almost absolutely failed in the final examination, and students who were successful to learning check testing more than two third out of all testing times took better score in the final examination.



\begin{figure}[htbp]
\begin{center}
\includegraphics[height=10cm]{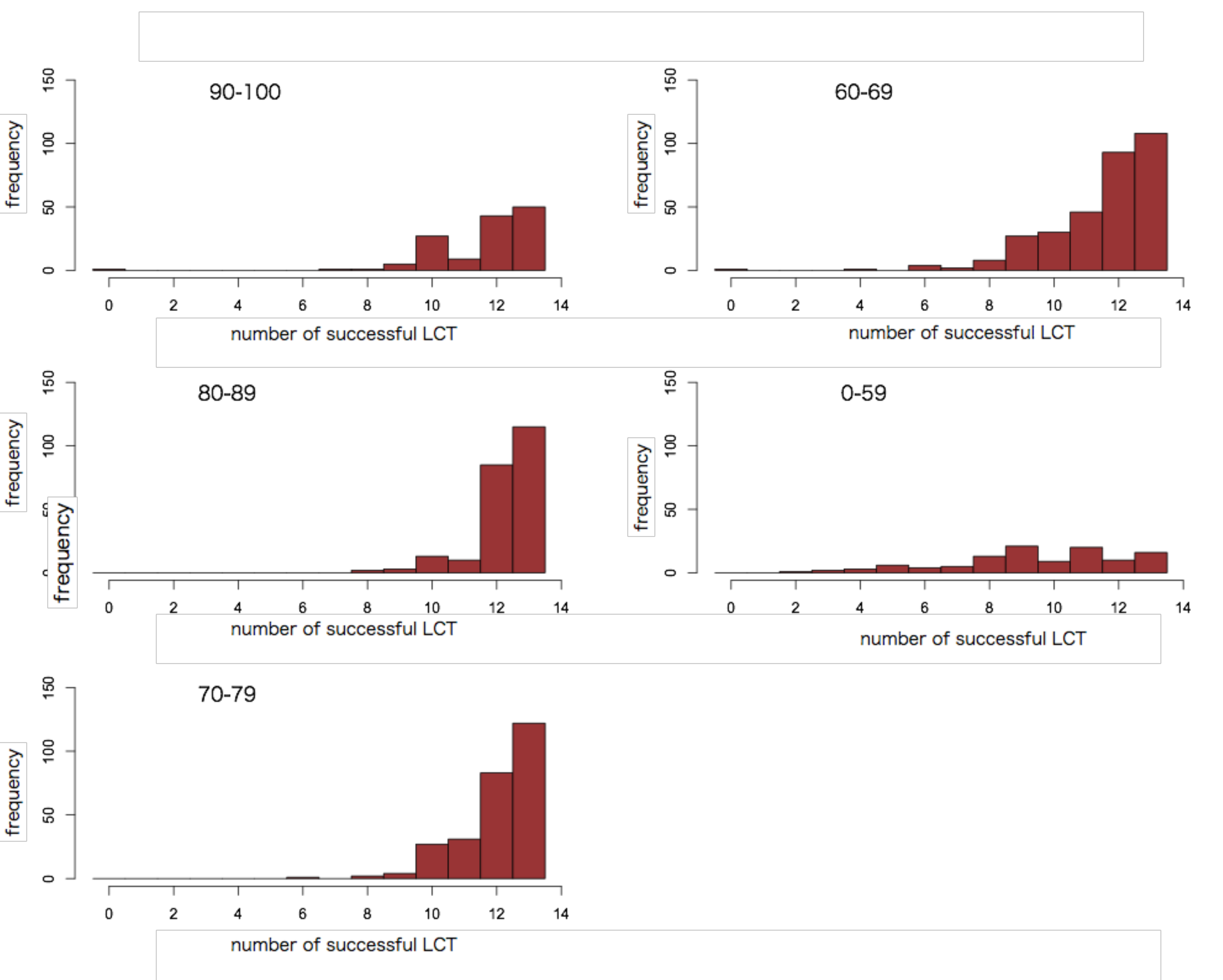}
\end{center}
\caption{Histograms of estimated abilities of LCT to successful students and to failed students (linear algebra in the first semester in 2017).}
\end{figure}

\section{Concluding Remarks}

It is crucial to identify students at risk for failing courses and/or dropping out as early as possible because a variety of students are now enrolled in universities and we teachers have to educate them altogether.
To overcome this, we established online testing systems aimed at helping students who need further learning skills in mathematics subjects,
including the learning check testing, the collaborative working testing, and the follow-up program testing.
Using the accumulated data from these testings in the database, 
we aimed at obtaining effective learning strategies for students at risk for failing courses and/or dropping out.
Although the subjects we deal with are analysis basic (similar to calculus) and linear algebra, we focused on  linear algebra case as a typical one.

In this paper, we have found some key factors strongly related to the students at risk. 
The findings are the following. 1) Badly failed students (score range is 0-39 in the final examination) tend to be  absent for the regular classes and fail in the learning check testing even if they attended, and they are very reluctant to attend the follow-up program classes. 2) Successful students (score range is 60-100 in the final examination) attend classes and get good scores in every learning check testing. 3) Failed students but not so badly (score range is 40-59 in the final examination) reveal both sides of features appeared in score range of 0-39 and score range of 60-100. Therefore, it is crucial to attend the lectures in order not to drop out.
Students who failed in learning check testing more than half out of all testing times almost absolutely failed in the final examination, which could cause the drop out. Also, students who were successful to learning check testing more than two third out of all testing times took better score in the final examination.

\section*{Acknowledgment}

The author would like to thank 
mathematical staffs at Hiroshima Institute of Technology.



\begin{thebibliography}{00}
\bibitem{Ayala} 
R. de Ayala, The Theory and Practice of Item Response Theory. Guilford Press, 2009.
\bibitem{Elouazizi} 
N. Elouazizi, Critical Factors in Data Governance for Learning Analytics, Journal of Learning Analytics, 1, 2014, pp. 211-222.
\bibitem{Siemens2015} 
D. Gasevic, S. Dawson, and G. Siemens, Let's not forget: Learning analytics are about learning, TechTrends, 59, 2015, pp. 64-71.
\bibitem{Hambleton91} 
R. Hambleton, H. Swaminathan, and H. J. Rogers, Fundamentals of Item Response Theory. Sage Publications, 1991.
\bibitem{LTLE2016a} 
H. Hirose, 
Meticulous Learning Follow-up Systems for Undergraduate Students Using the Online Item Response Theory, 5th International Conference on Learning Technologies and Learning Environments, 2016, pp.427-432.
\bibitem{LTLE2016b} 
H. Hirose, M. Takatou, Y. Yamauchi, T. Taniguchi, T. Honda, F. Kubo, M. Imaoka, T. Koyama, 
Questions and Answers Database Construction for Adaptive Online IRT Testing Systems: Analysis Course and Linear Algebra Course, 5th International Conference on Learning Technologies and Learning Environments, 2016, pp.433-438.
\bibitem{LTLE2016c} 
H. Hirose, 
Learning Analytics to Adaptive Online IRT Testing Systems ``Ai Arutte'' Harmonized with University Textbooks, 5th International Conference on Learning Technologies and Learning Environments, 2016, pp.439-444.
\bibitem{LTLE2017} 
H. Hirose, M. Takatou, Y. Yamauchi, T. Taniguchi, F. Kubo, M. Imaoka, T. Koyama, 
Rediscovery of Initial Habituation Importance Learned from Analytics of Learning Check Testing in Mathematics for Undergraduate Students, 6th International Conference on Learning Technologies and Learning Environments, 2017, pp.482-486.
\bibitem{BIC2016} 
H. Hirose, Dually Adaptive Online IRT Testing System, Bulletin of Informatics and Cybernetics Research Association of Statistical Sciences, 48, 2016, pp.1-17.
\bibitem{IEE2018} 
H. Hirose, Difference Between Successful and Failed Students Learned from Analytics of Weekly Learning Check Testing, Information Engineering Express, Vol 4, No 1, 2018, pp.11-21.
\bibitem{PISM2018} 
H. Hirose, A Large Scale Testing System for Learning Assistance and Its Learning Analytics, Proceedings of the Institute of Statistical Mathematics, Vol.66, No.1, 2018, pp.79-96.
\bibitem{LindenHambleton} 
W. J. D. Linden and R. K. Hambleton, Handbook of Modern Item Response Theory. Springer, 1996.
\bibitem{IJSCAI2017} 
T. Sakumura, H. Hirose, Bias Reduction of Abilities for Adaptive Online IRT Testing Systems, International Journal of Smart Computing and Artificial Intelligence (IJSCAI), 1, 2017, pp.57-70.
\bibitem{Siemens2012}
G. Siemens and D. Gasevic, Guest Editorial - Learning and Knowledge Analytics, Educational Technology \& Society, 15, 2012, pp.1-2.
\bibitem{LTLE2016d} 
Y. Tokusada, H. Hirose, 
Evaluation of Abilities by Grouping for Small IRT Testing Systems, 5th International Conference on Learning Technologies and Learning Environments, 2016, pp.445-449.
\bibitem{Waddington2016} 
R. J. Waddington, S. Nam, S. Lonn, S.D. Teasley, ,  Improving Early Warning Systems with Categorized Course Resource Usage, Journal of Learning Analytics, 3, 2016, 263-290.
\bibitem{WiseShaffer} 
A.F. Wise and D.W. Shaffer, Why Theory Matters More than Ever in the Age of Big Data, Journal of Learning Analytics, 2, pp. 5-13, 2015.
\bibitem{R} 
\url{https://www.r-project.org/about.html}
\end{thebibliography}
\end{document}